\documentclass[]{emulateapj}
\usepackage{epsf}

\def\ltsima{$\; \buildrel < \over \sim \;$}
\def\lsim{\lower.5ex\hbox{\ltsima}}
\def\Msunh{\mbox{$h^{-1}$M$_\odot$}}
\def\mpch{\mbox{$h^{-1}$Mpc}}

\def\deg{\ifmmode{^\circ}\else{$^\circ$}\fi}
\def\hGpc{\ifmmode{h^{-1}{\rm Gpc}}\else{$h^{-1}{\rm Gpc}$}\fi}
\def\hkpc{\ifmmode{h^{-1}{\rm kpc}}\else{$h^{-1}{\rm kpc}$}\fi}
\def\hMpc{\ifmmode{h^{-1}{\rm Mpc}}\else{$h^{-1}{\rm Mpc}$}\fi}
\def\hMsun{\ifmmode{h^{-1}M_\odot}\else{$h^{-1}M_\odot$}\fi}

\def\muK{\ifmmode{\mu{\rm{K}}}\else{$\mu$K}\fi}
\def\mum{\ifmmode{\mu{\rm{m}}}\else{$\mu$m}\fi}

\newcommand{\LCDM}{$\Lambda$CDM}

\begin{document}

\title{The growth of galaxy stellar mass within dark matter halos}

\author{Idit Zehavi\altaffilmark{1}, Santiago Patiri\altaffilmark{1,2}, and
Zheng Zheng\altaffilmark{3}}

\altaffiltext{1}{Department of Astronomy \& CERCA, Case Western Reserve
University,  10900 Euclid Avenue, Cleveland, OH 44106}
\altaffiltext{2}{IANIGLA-CONICET, C.C. 330, 5500 Mendoza, Argentina}
\altaffiltext{3}{Yale Center for Astronomy and Astrophysics, Department of Physics, Yale University, New Haven, CT 06520}

\begin{abstract}

We study the evolution of stellar mass in galaxies as a function of host 
halo mass, using the ``MPA'' and ``Durham'' semi-analytic models, implemented 
on the Millennium Run simulation. The results from both models are similar.
We find that about $45\%$ of the stellar mass in central galaxies in 
present-day halos less massive than $\sim10^{12}\hMsun$ is already in place 
at $z\sim1$. This fraction increases to $\sim65\%$ for more massive halos. The 
peak of star formation efficiency shifts toward lower mass halos from $z\sim1$
to $z\sim0$. The stellar mass in low-mass halos grows mostly by star formation 
since $z\sim1$, while in high-mass halos most of the stellar mass is assembled 
by mergers. These trends are clear indications of ``halo downsizing''. We 
compare our findings to the results of the phenomenological method developed 
by Zheng, Coil \& Zehavi (2007). The theoretical predictions are in qualitative
agreement with these results, however there are large discrepancies. The most 
significant one concerns the amount of stars already in place in the progenitor
galaxies at $z\sim1$, which is about a factor of two larger in both 
semi-analytic models. We also use the semi-analytic catalogs to test different 
assumptions made in that work, and illustrate the importance of smooth 
accretion of dark matter when estimating the mergers contribution. We 
demonstrate that methods studying galaxy evolution from the galaxy-halo 
connection are powerful in constraining theoretical models and can guide future
efforts of modeling galaxy evolution. Conversely, semi-analytic models serve 
an important role in improving such methods. 

\end{abstract}


\section{Introduction}

In the current paradigm of structure formation, galaxies form within cold 
dark matter halos. The formation and evolution of these halos are 
dominated by gravity and can be well predicted from high-resolution 
cosmological numerical
simulations and analytic models. The assembly of the stellar content of
galaxies, however, is governed by more complex physics, and
the relation between galaxies and dark matter halos and
the detailed physical processes of galaxy formation and evolution 
are  only partially understood. 

A useful approach to explore galaxy formation within 
dark matter halos is the Semi-Analytic Modeling (SAM) of galaxy formation 
(e.g., \citealt{Cole94, Cole00, Benson03, Croton06, Bower06}). In such
models, 
halos identified from high resolution $N$-body simulations are ``populated'' 
with galaxies using analytical prescriptions for the baryonic 
evolution. Within the SAM approach, galaxies change as the original stars 
evolve and new stars form. They also change their stellar content and 
increase their mass by merging with other galaxies. Different feedback or 
pre-heating mechanisms, such as those caused by star formation, active 
galactic nuclei, or the photo-ionizing ultra-violet background, also impact 
at different stages of a galaxy's life and are implemented in the models 
at different levels. These models have been successful in reproducing 
several measured properties including the galaxy luminosity and stellar 
mass functions (see e.g., \citealt{Croton06, Bower06, Fontanot09}).

Different phenomenological methods 
have been developed to connect galaxies with dark matter halos. 
One commonly used approach is the Halo Occupation Distribution framework
(HOD, e.g., \citealt{Jing98, Peacock00, Seljak00, Scoccimarro01, Berlind02}),
which characterizes the relationship between galaxies 
and halos in terms of the probability distribution, $P(N|M)$, that a halo 
of virial mass $M$ contains $N$ galaxies of a given type, together with 
the spatial and velocity distributions of galaxies inside halos. The HOD 
parameters are constrained using galaxy clustering measurements from large 
galaxy surveys and theoretically known halo clustering. 
Similar approaches include the Conditional Luminosity Function 
(CLF, see \citealt{Yang03}), which describes the average number of 
galaxies as a function of luminosity that reside in a halo of mass 
$M$, and abundance matching schemes 
(\citealt{Conroy06,Conroy09,Moster09,Guo10,Neistein11,Rodriguez11}),
which monotonically connect galaxy luminosity or stellar mass to halo mass 
by matching the abundances of halos and galaxies.

HOD models have been mostly used to learn about the relationship between 
galaxies and halos at a fixed epoch (e.g., \citealt{Bullock02, Zehavi05, 
Zehavi10, Zheng08} and references therein). 
Recent studies have started using them to also explore galaxy evolution
by combining the inferred galaxy-halo 
connection at different redshifts with the evolution of dark matter halos 
provided by theory \citep{ZCZ07,White07,Seo08,Wake08,Wake10}. 
In particular,  Zheng, Coil \& Zehavi (2007; hereafter ZCZ07) develop 
a phenomenological approach to extract information 
about galaxy evolution from $z\sim 1$ to $z\sim 0$ by performing HOD 
modeling of two-point correlation functions of DEEP2 and SDSS galaxies. 
With the inferred galaxy-halo connection at two redshifts, they establish 
an evolutionary link using the typical growth of dark matter halos
obtained from numerical simulations. 

Even with the progress made in establishing the evolutionary link between 
galaxies and halos, our understanding of the specifics of stellar mass
growth within the dark matter halos is still far from complete. 
Galaxies can grow their stellar mass by star formation, accretion of smaller 
satellite galaxies or major merging. It is important to 
quantify the contribution of all these processes in order to have a complete 
picture of the assembly history of galaxies within their host dark matter 
halos. ZCZ07 derive the mean stellar masses of central galaxies at
$z\sim1$ and $z\sim0$ as a function of the present-day host halo mass.
After roughly accounting for the contribution of merging of central and
satellite galaxies, they infer the star-formation contribution to the
stellar mass assembly.
They find that in central galaxies located in relatively low-mass halos 
($\sim 5\times 10^{11} \Msunh$) the bulk of the stellar mass 
results from star formation between $z\sim1$ and $z\sim 0$, while only
a small fraction of stars formed since $z\sim1$ in central galaxies of
halos as massive as $\sim 5\times 10^{12} \Msunh$. For these massive halos,
merging becomes more important and constitutes the dominant contribution 
to the stellar mass growth 
(see Fig. 9 in ZCZ07 for details). The results reflect the so-called
``downsizing'' star formation pattern in which the sites of active star
formation shift from high-mass galaxies at early times to lower-mass
systems at later epochs (e.g., \citealt{Cowie96}), manifested in terms of
halo mass.
 
In this paper, we study the theoretical predictions for stellar mass evolution 
as a function of dark matter halo mass using SAM catalogs. 
One of the main advantages of the SAMs
is that we can trace the full evolution of the individual galaxies within 
their dark matter halos, allowing an explicit study of the different 
processes that contribute to the build up of the stellar content of galaxies. 
We compare and contrast these predictions with the results of ZCZ07.
We gauge the potential of the phenomenological methods to constrain galaxy 
formation models, as well as test some of the assumptions of such methods. 
In particular, we check the validity of the 
simple evolutionary approach presented by ZCZ07. 
 
For these purposes, we use two SAM catalogs, the ``MPA'' 
(\citealt{Croton06, DeLucia07}) and ``Durham'' (\citealt{Bower06, font08}) 
catalogs produced from the Millennium Run cosmological {\it N}-body simulation 
(\citealt{Springel05}). We mainly focus on the stellar mass growth as a 
function of halo mass since $z\sim 1$ for ease of comparison with ZCZ07.
Nonetheless, we briefly study also the stellar mass growth since $z\sim 2$ 
in the MPA catalog to explore the processes involved in galaxy 
formation at higher redshifts.
These SAM catalogs have been previously used to study the stellar 
mass evolution in galaxies (e.g., \citealt{Guo08,Stringer09,Fontanot09}).
However, those studies focus on the integrated stellar mass to 
make a direct comparison to observations and did not investigate in detail 
its evolution as a function of halo mass. Most physical processes 
involved in galaxy formation models depend explicitly on the mass of the 
dark matter halo. Additionally, the modeling of other observables such 
as the galaxy-galaxy merger rate strongly rely on the precise knowledge of 
the galaxy-halo connection (see e.g., \citealt{Hopkins10}). Thus it is 
physically meaningful and informative to study galaxy evolution as a 
function of halo mass. 

The paper is organized as follows: \S2 describe the SAM catalogs that
we use and the sample selection. In \S3 we present and discuss 
our results for the growth of stellar mass as a function of halo mass.
In \S4 we compare our results to those found by ZCZ07 and we conclude
in \S5.

\section{Mock Catalogs and Galaxy Formation Models}
\label{sec:samples}

In this work, we use the publicly available mock galaxy catalogs produced 
with the ``MPA'' (\citealt{Croton06, DeLucia07}) and ``Durham'' 
(\citealt{Bower06}) semi-analytic models of galaxy formation, both based 
upon dark matter halo evolution in the Millennium Run simulation 
(\citealt{Springel05}). The Millennium Run followed the evolution of 
$\sim 10^{10}$ dark matter particles in a \LCDM~ Universe. The simulation uses
a periodic box of $500 \mpch$ on a side with mass resolution per 
particle of $8.6 \times 10^8 \Msunh$. The initial conditions of the 
simulation were generated with cosmological parameters obtained from the 
combined analysis of 2dFGRS and WMAP1 CMB data. The halos in the simulation 
were identified in each time step using a friend-of-friends algorithm with a 
linking length of 0.2 times the mean particle separation. More details 
can be found in \citealt{Springel05}.

In the SAMs, galaxies are assumed to form at the center of dark matter 
halos. The evolution of the baryonic component of galaxies is modeled using 
simple but physically motivated analytic prescriptions. These include 
radiative cooling of hot gas, star formation in the cold disk, supernova 
feedback, black hole growth and AGN feedback through the ``quasar'' and 
``radio'' epochs of AGN evolution, metal enrichment of the inter-galactic 
and intra-cluster medium, as well as galaxy morphology shaped through 
mergers and merger-induced starbursts.  

As mentioned above, these models are aimed at reproducing 
integrated galaxy observables. To that effect, the SAMs recover 
reasonably well the galaxy luminosity function in different bands (e.g., 
\citealt{Croton06, Bower06}) as well as the stellar mass functions for a 
range of redshifts (e.g., \citealt{Bower06, Fontanot09}; but see also 
\citealt{li09,Marchesini09}). The 
predictions for the fraction of blue and red central galaxies, however, are 
not fully 
correct (see \citealt{Baldry06}), and modifications to the treatment of gas 
cooling (e.g., \citealt{Viola08}) and the AGN feedback (e.g., 
\citealt{Baldry06}) have been proposed to alleviate these discrepancies. 
The SAMs also appear to overestimate the fraction of red satellites 
(\citealt{Weinmann06,Coil08, Kang08, DeLucia09, SeekKim09}). Note that 
satellite 
galaxies in these models are treated differently than central galaxies. 
Only central galaxies accrete new material by cooling from the hot 
atmosphere of their halo, direct infall of cold gas, and the merging of 
satellites. Since no new material accretes onto satellites, their star 
formation ends when the cold gas is exhausted (see \citealt{Croton06}).  
Satellites are also affected by different environmental processes that
change their properties, which are not fully 
implemented in these models. For example, \citet{SeekKim09} suggest that 
satellite-satellite mergers and tidal dissolution of satellites need to be 
included to better match the measured luminosity 
dependence of galaxy clustering. 

Even though different SAMs adopt similar analytical prescriptions to treat 
the physical processes involved in galaxy formation and evolution, there are 
significant differences in the way the MPA and Durham SAMs deal with specific  
processes, such as the cooling of gas, the cut-off black hole mass for AGN 
feedback (see \citealt{Stringer09}) and the dynamical treatment of the 
``orphan'' 
satellite galaxies (\citealt{Gao04}). Also, the two models are based on 
different halo merger trees which are constructed in the post-processing 
stage of the simulation (see \citealt{Bower06, DeLucia07}). Even though 
the halo merger trees are checked to be statistically compatible (De Lucia, 
private communication), differences could arise at the level of individual 
galaxies, affecting some galaxy properties. 

The full SAM galaxy catalogs used in this 
work\footnote{The SAM galaxy catalogs can be downloaded 
from http://www.g-vo.org/Millennium/}
contain information for about eight
million galaxies brighter than $M_r=-17$. Several properties are available
for each of these galaxies, including positions and velocities, magnitudes 
in several band passes (Johnson, Busher, 2MASS as well as the 5 SDSS bands), 
stellar mass, and mass of the dark matter halo in which the galaxy is located. 
It is important to note that these two SAMs assume different initial mass 
functions  of stars. The MPA model assumes a Chabrier mass function 
\citep{Chabrier03}, while the Durham model uses the Kennicutt one
\citep{Kenicutt83}. In order to make a direct comparison between the models, 
we consistently transform both to a ``diet'' Salpeter initial mass function 
\citep{Bell03}, used also in ZCZ07. 

We select from each catalog a random sample of 250,000 present-day central 
galaxies.  For compatibility with the ZCZ07 results,
we use in fact the $z \sim 0.1$ snapshot of the catalogs, as  
this is approximately the value of the median redshift for SDSS galaxies.
We hereafter, however, loosely refer to it as $z\sim0$, in comparison to
the evolution since $z\sim1$ and $z\sim2$.
For each present-day galaxy, we identify all its 
progenitors at $z\sim 1$ (and also at $z\sim 2$ in the MPA SAM) by following 
the merger tree of each present-day central galaxy. 
These progenitors can be central galaxies in dark matter halos, satellites 
located in subhalos or ``orphan'' galaxies, i.e. satellite 
galaxies whose parent dark matter subhalo was destroyed below the resolution 
limit of the simulation by tidal stripping and truncation (see, e.g., 
\citealt{Gao04}). We define the {\it main} progenitor of a 
$z \sim 0$ galaxy
as the central galaxy located in the most massive dark matter halo at the 
corresponding higher redshift ($z\sim 1$ for both catalogs and $z\sim 2$ for 
the MPA only). We have verified that defining the main progenitor as the 
most massive of the merger tree main branch (e.g., \citealt{DeLucia07}) 
does not change 
our results. Once all progenitors are identified, we can
quantify the different contributions to the stellar mass of central 
galaxies, namely the contributions from smaller central galaxies and from
satellites that merge into the central galaxies. The present-day stellar mass 
in central galaxies that does not come from the different merger processes 
arises then  from  new star formation during that time period.

\section{Stellar Mass Growth as a Function of Halo Mass}

In this section we first evaluate the star formation efficiency of galaxies 
as a function of host halo mass.  We then study the evolution of stellar 
mass in central galaxies in the semi-analytic models. 
We also analyze the different contributions to the stellar mass growth of 
central galaxies since $z \sim 1$. Finally, we briefly investigate  
the stellar mass growth since $z \sim 2$.

\subsection{Star Formation Efficiency}

A first fundamental quantity to investigate is the star formation 
efficiency (SFE) as a function of halo mass. Here, we define the SFE as 
the total stellar mass in central galaxies ($M_{\rm star}$) divided by the 
baryon mass $M_{b} = f_{b}M$, 
where $f_{b}$ is the global baryon fraction and $M$ is the mass of the 
dark matter halo. This SFE represents the integrated value from 
the redshift of formation to the epoch in consideration, reflecting the 
fraction of baryons associated with halos that are converted into stars 
in that time period.

\begin{figure}
\includegraphics[width=\columnwidth]{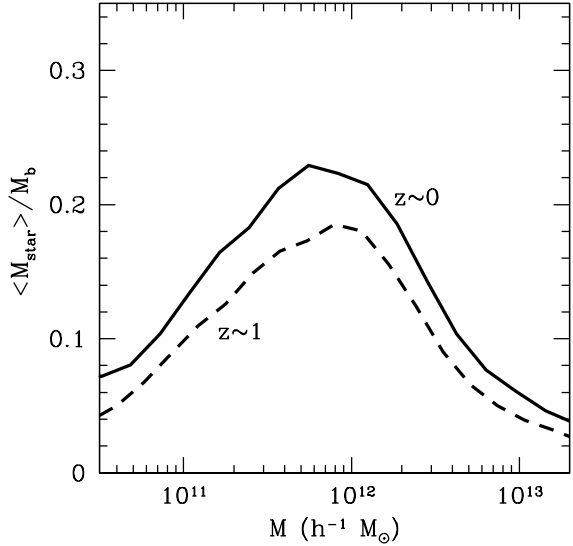}
\caption{Star formation efficiency at $z \sim 0$ (solid line) and at 
$z \sim 1$ (dashed line) as a function of halo mass for the MPA SAM.
} 
\label{fig:Fig10mill50b}
\end{figure}

Figure~\ref{fig:Fig10mill50b} shows the SFE at $z \sim 1$ and 
$z \sim 0$ as a function of halo mass predicted by the MPA SAM. 
The baryon fraction adopted in the MPA SAM is $f_b=0.17$, though the baryon
fraction in any individual halo may vary from the global value.
The SFE at both epochs has a similar shape, peaking at some characteristic 
mass and dropping towards both low and high mass ends.
At $z\sim 0$, the SFE peaks for halos of mass $\sim 5\times 10^{11} \Msunh$, 
while at $z\sim 1$ the peak
corresponds to  more massive halos ($\sim 10^{12} \Msunh$). 
This trend is an indication of the ``downsizing'' phenomena 
(e.g., \citealt{Cowie96,Juneau05,Fontanot09,Avila11}), shown specifically
here as a function of halo mass. The SFE at the peak is $\sim 23\%$ 
and $\sim 18\%$ at $z \sim 0$ and $z \sim 1$, respectively.
The SFE results for the Durham model are similar. 

The drop of SFE at the low-mass end is associated
with the availability of cold gas in these halos, which can be affected by
photo-ionization heating and star formation feedback.
The drop of SFE above the characteristic halo mass is related to
gas accretion becoming less efficient due to the high virial temperature
and AGN feedback. Additionally, we are only considering here central
galaxies, and in high-mass halos the stellar mass contributions from 
satellite galaxies can be substantial.

\subsection{Stellar Mass Growth of Central Galaxies}

\begin{figure}
\includegraphics[width=\columnwidth]{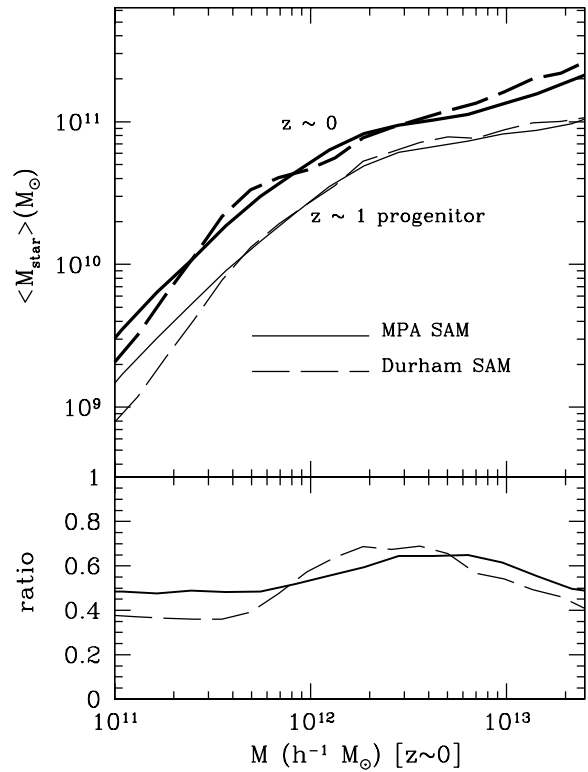}
\caption{
Mean stellar mass in $z\sim 0$ central galaxies and their $z\sim 1$ progenitor
central galaxies as a function of the present-day halo mass.
{\it Top panel}: The mean stellar mass in central galaxies at $z 
\sim 0$ (thick lines) and that of their $z \sim 1$ main progenitors (thin 
lines) as a function of the present-day halo mass, predicted by the MPA 
(solid lines) and Durham (dashed lines) SAMs. {\it Bottom panel}: The ratio 
between the stellar mass at $z\sim 1$ and $z\sim 0$ for the MPA (solid 
line) and Durham (dashed line) SAMs, which represents the fraction of 
$z\sim 0$ central stellar mass that is in place in the $z\sim 1$
progenitor central galaxies.
}
\label{fig:Fig8MPADurham}
\end{figure}

Figure \ref{fig:Fig8MPADurham} shows the mean stellar mass in central 
galaxies at $z \sim 0$ (thick lines) and that of their $z \sim 1$ main 
progenitors (thin lines) as a function of the present-day halo mass. The
results for the MPA catalog (solid lines) and the Durham catalog
(dashed lines) are similar. 
At both redshifts, the stellar mass of the central galaxy 
increases rapidly with halo mass at the low mass end and slowly at the high 
mass end. The transition halo mass is approximately $10^{12} \Msunh$ at
$z \sim 0$ and $2 \times 10^{12} \Msunh$ at $z \sim 1$. 

The bottom panel shows the ratio between the stellar mass at $z \sim 1$ and 
$z \sim 0$, representing  the fraction of the $z\sim0$ central galaxy stellar 
mass already in place in the progenitor central galaxies at $z\sim 1$,
as a function of present-day halo mass. 
For example, for a present-day halo of mass $5 \times 10^{11} \Msunh$, on 
average $\sim 50\%$ ($\sim 40\%$) of the stellar mass in the central galaxy 
is already in place in the $z\sim 1$ main progenitor central galaxy for the 
MPA (Durham) model. The ratio gradually increases to $\sim 65\%$ for halos 
with mass $\sim$ a few $\times 10^{12} \Msunh$, and it starts 
decreasing towards the highest halo masses probed in this work. 
The decrease of stellar mass ratio at the highest halo masses is in accord 
with the predictions for hierarchical assembly of massive galaxies (e.g., 
\citealt{DeLucia06, DeLucia07}).

\subsection{Different Contributions to the Stellar Mass Growth}

The stellar mass in galaxies grows as a consequence of internal star 
formation or external infall of material (major and minor mergers). For the 
former, observational estimates (e.g., \citealt{Salim07, Noeske07}) 
often come from measuring the average star formation rate as a function of 
stellar mass and time. The amount of stellar mass gained through accretion 
is more difficult to measure directly. It is often estimated by simply 
taking the difference between the growth due to star formation and the total 
stellar mass at present. 

In a SAM based on an $N$-body cosmological simulation, we 
have the full merger history of dark matter halos and galaxies. Thus, it 
is possible to track the complete evolution of the stellar mass due to both
mergers and star formation as a function of the halo mass. In particular, 
following ZCZ07, we account for four different components of the assembly of 
stellar mass in central galaxies: stars in place in the progenitor central 
galaxies at $z\sim 1$, stellar mass from smaller central galaxies that merge 
to the 
central galaxy, stellar mass from any satellites (of the main progenitor or 
other smaller central galaxies) that merge with the central galaxy,
and recent star formation.

\begin{figure*}
\includegraphics[width=165mm]{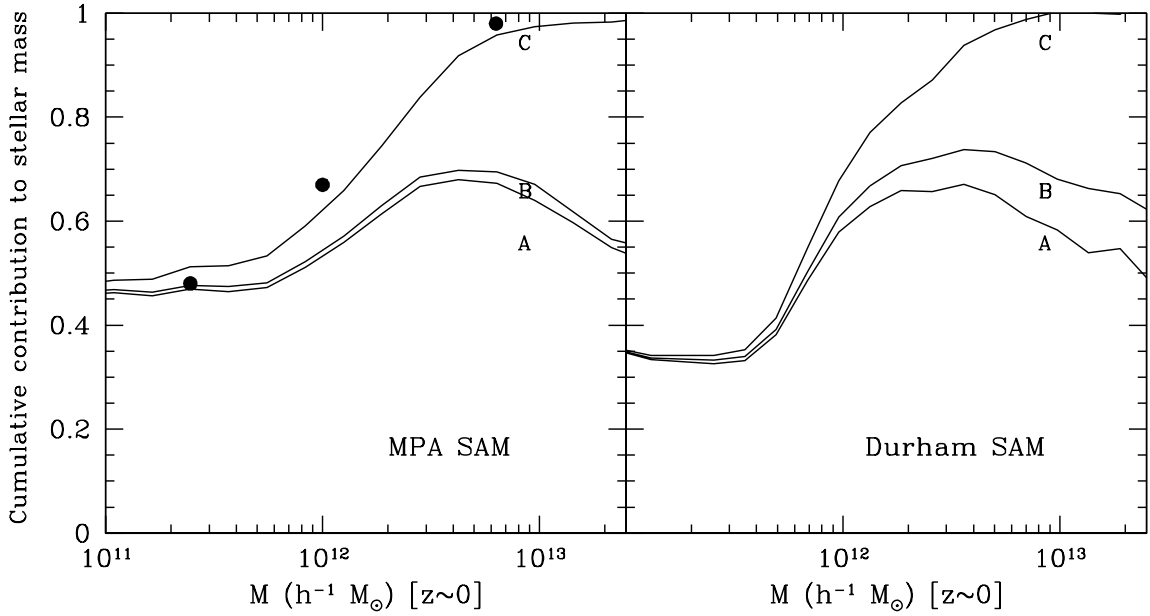}
\caption{Different contributions to the present-day central galaxy stellar 
mass as a function of present-day halo mass shown for the MPA (left panel) 
and Durham (right panel) SAMs. The first curve from the bottom (denoted as 
``A'') is the stellar mass already in place in the progenitor central galaxies 
at $z\sim 1$. The area between lines ``A'' and ``B'' indicates the 
contribution of the smaller central galaxies that merge with the progenitor 
central galaxies. The area between curve ``B'' 
and ``C'' denotes the contribution of satellite galaxies. The remainder 
(from curve ``C'' to the top of the plot) is the stellar mass gained through 
star formation since $z\sim1$. The full circles in the left panel are the 
contribution of star formation in the main progenitors obtained directly 
by integrating the star-formation rate over time (see text for details).}
\label{fig:Fig9MPADurham}
\end{figure*}

Figure \ref{fig:Fig9MPADurham} presents the SAM predictions for these
different contributions to the $z \sim 0$ central galaxy stellar mass as a 
function of the present-day halo mass, for both the MPA and Durham models. 
In each panel, the bottom curve (marked as ``A'') denotes the fraction of 
stellar mass in place in the $z\sim 1$ progenitor central galaxies and is essentially the same 
ratio shown in the bottom panel of Figure \ref{fig:Fig8MPADurham}. 
The area between curves ``A'' and ``B'' indicates the contribution 
of smaller central galaxies that have merged into the main progenitor. The 
area between curves ``B'' and ``C'' shows the contribution from satellite 
galaxies in all progenitor halos at $z\sim 1$ that end up in the $z\sim 0$ 
central 
galaxies. Consequently, the remaining stellar mass (from curve 
``C'' to the top of the plot) arises from star formation since $z \sim 1$.

In the SAM datasets, the amount of star formation since $z\sim 1$ can also 
be directly obtained by integrating the star formation rate as a function 
of time. We do a crude calculation of this in the MPA SAM for three specific 
present-day halo masses, as a sanity check, by integrating the star-formation 
rate for a subsample of main progenitors over 10 snapshots since 
$z \sim 1$ (shown as 
filled circles in the left panel of Figure~\ref{fig:Fig9MPADurham}).
This gives a sense of the total star formation since $z \sim 1$, since 
the contribution from smaller central progenitors is minor and the
satellites do not contribute to the star formation in these models
(see \S~2).  The first point appears to be a bit below the
expected value (corresponding to a larger star formation contribution),
reflecting the approximate nature of this calculation.

The overall trends of stellar mass growth from the two SAMs are similar.
It is evident that central galaxies in small halos grow mostly by star 
formation 
since $z \sim 1$, while the star formation contribution is small in high 
mass halos. This could be explained by the fact that small halos are almost 
completely assembled by $z \sim 1$, so the new stellar mass must come from 
star formation. For intermediate halo masses, the contribution from merging 
becomes more significant, but star formation still contributes most of 
the stellar mass since $z \sim 1$. For high-mass halos, the contribution 
from mergers dominates and star formation is negligible. This overall behavior 
is another manifestation of the ``downsizing'' effect, in this case 
referring to the fact that more massive galaxies form the bulk of their 
stars earlier than smaller galaxies (``anti-hierarchical'' 
assembly; see e.g., \citealt{DeLucia06, Stringer09, Fontanot09}). 

Some differences between the two SAM models are apparent.
The stellar mass already in place in the $z\sim1$ progenitors
was compared in the bottom panel of Figure~\ref{fig:Fig8MPADurham}. With
regard to the other components, the Durham model predicts a larger 
contribution from smaller central progenitors in halos with present-day 
mass larger than $\sim 10^{12} \Msunh$, while the MPA model produces a 
slightly larger contribution from satellite galaxies in low-mass halos.
These discrepancies likely reflect the differences in the galaxy formation 
prescription in the two models as well as the differences in the underlying
halo merger trees. Another contributing factor is the different treatment
of the dynamics of ``orphan'' galaxies (De Lucia, private communication).

\subsection{Stellar Mass Growth Since $z \sim 2$}

In this section we explore the SAM predictions for stellar mass growth from 
higher redshifts, to gain further insight on the processes contributing to
galaxy evolution and serve as a reference point for modeling higher-redshift 
observations. The interpretation of these predictions is complex since 
observations currently do not provide consistent indicators of stellar 
masses and star formation rates at these redshifts (see e.g., 
\citealt{Conroy09} \S3.5 for a discussion). 
Also clustering data at $z \sim 2$ is relatively scarce for ZCZ07-like 
analyses (but see \citealt{Wake10} for a first attempt
along these lines). 

\begin{figure}
\includegraphics[width=\columnwidth]{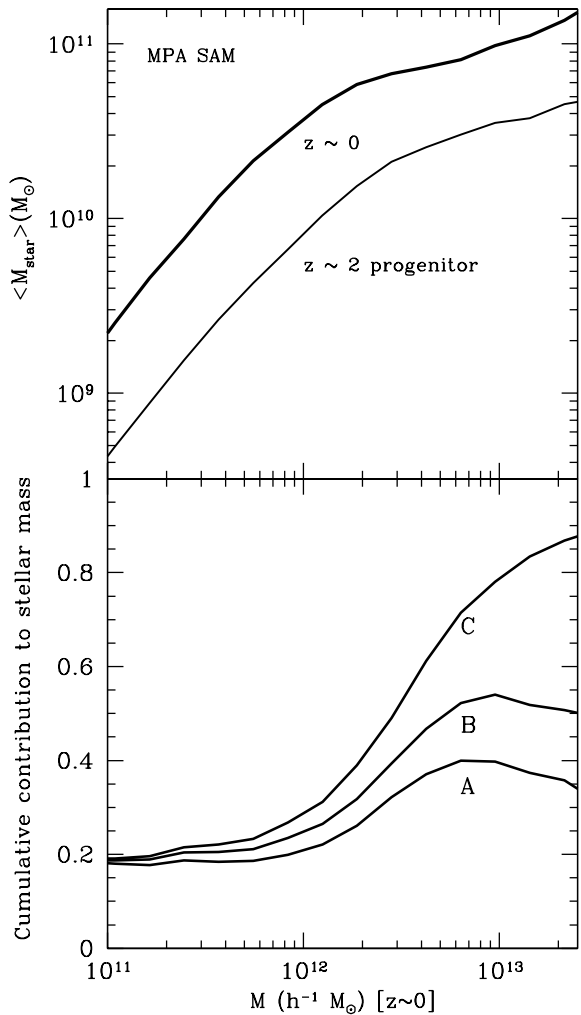}
\caption{{\it Top panel}: Mean stellar masses in central galaxies at $z \sim 
0$ (thick line) and that of their $z \sim 2$ main progenitors (thin 
line), as a function of the present-day halo mass, predicted by the MPA SAM. 
{\it Bottom panel}: The different contributions to 
the present-day central galaxy stellar mass. Curves are the same as 
in Fig.~\ref{fig:Fig9MPADurham}, except now referring to stellar mass evolution
since $z\sim2$.}
\label{fig:Fig9MPAz2}
\end{figure}

We study the stellar mass growth as a function of halo mass since 
$z\sim 2$ in the MPA catalog, using a smaller sample of $160,000$ 
galaxies. 
The SFE at $z\sim2$ as a function of halo mass exhibits a similar shape
and peak location as the $z\sim1$ results (shown in 
Fig.~\ref{fig:Fig10mill50b}). The overall amplitude is slightly smaller, 
with a SFE of $\sim 16\%$ at the peak of the distribution. 
Our results for the growth of stellar mass are presented in 
Figure~\ref{fig:Fig9MPAz2}. The top panel shows the mean stellar mass 
in central galaxies at $z\sim0$ (thick line) and that of their $z \sim 2$ 
main progenitors (thin line), as a function of the present-day halo mass. 
The bottom panel presents the different contributions to
the present-day central galaxy stellar mass. 
The different curves in this plot are the same as those in 
Figure~\ref{fig:Fig9MPADurham}, but referring now to $z\sim 2$, with curve
``A'' representing the stars already in place in the central galaxies at 
$z\sim 2$. 

The trends are qualitatively very similar to the ones seen for the 
$z\sim1$ progenitors in the previous sections. 
As expected, there is significantly less stellar mass already in place in the 
main progenitor at $z\sim 2$. 
The contribution from the smaller central galaxies is more significant 
than that for the $z\sim 1$ case, while the contribution of the satellites 
is roughly the same. As less stellar mass is in place at $z\sim2$,  the 
contribution of star formation from $z\sim2$ to the present-day is 
substantial, at all halo masses; it is about $10\%$ for central galaxies 
in the highest mass halos probed and higher for those in lower mass halos. 
Note that the ``downsizing'' pattern is also evident here.

\section{Comparison with Phenomenological Approach}

Zheng, Coil, \& Zehavi (2007; ZCZ07) perform HOD modeling of the 
luminosity-dependent projected two-point correlation function for 
DEEP2 and SDSS galaxies, at $z\sim1$ and $z\sim0$, respectively. They infer 
the relationship between central 
galaxy luminosity and halo mass at these two redshifts and establish an 
evolutionary link by using the typical growth of dark matter halos obtained 
from numerical simulations. Stellar masses are derived from the galaxies 
luminosity and color. As a proof of concept, they estimate the evolution 
of galaxy stellar mass as a function of host halo mass. 
An approximate method is used to estimate the different contributions 
of mergers and star formation to the growth of central galaxies stellar mass.
In this section, we use the SAM results to gauge the potential of such
phenomenological methods to constrain galaxy formation and evolution models.
We first compare the stellar
mass growth in central galaxies inferred by ZCZ07 from DEEP2 and SDSS galaxy 
clustering with that predicted by the SAM models. We then examine the 
validity of the assumptions used in the ZCZ07 approach.

\subsection{Star Formation Efficiency and Galaxy Stellar Mass Growth} 

We first examine the star formation efficiency and its evolution from
redshift $z\sim1$ to $0$ obtained using the MPA SAM model (our 
Fig.~\ref{fig:Fig10mill50b}) and in ZCZ07 (their Fig.~10). Both approaches
produce similar general trends with halo mass, exhibiting peaked 
distributions at the two redshifts, and at similar halo masses. The ZCZ07 
results also exhibit the halo ``downsizing'' effect, with the SFE peak
shifting to a higher mass at the higher redshift. The MPA SAM and 
ZCZ07 also find comparable values for the maximum SFE at $z \sim 0$. 
However, at $z \sim 1$, the SAM shows an overall higher SFE than computed 
in ZCZ07 (peaking at $18\%$ and $12\%$, respectively), 
which translates into a larger amount of stars by that redshift. 

\begin{figure*}
\includegraphics[width=165mm]{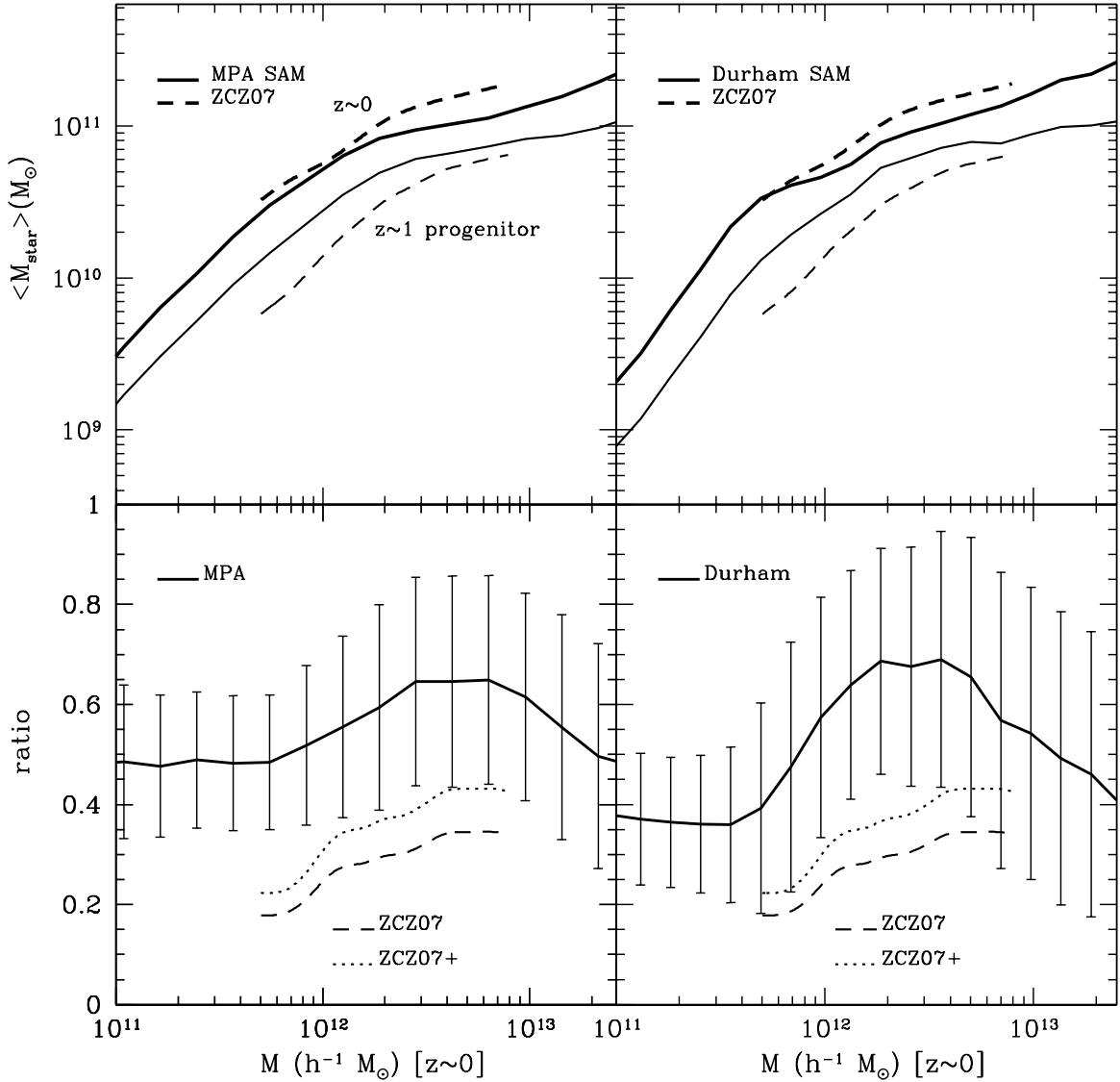}
\caption{{\it Top panels}: Comparison of the MPA (left, solid lines) 
and Durham (right, solid lines) SAMs with the ZCZ07 results (dashed lines) 
for the mean stellar mass in central galaxies at $z\sim 0$ and $z\sim 1$, as a 
function of present-day halo mass. {\it Bottom panels}: Comparison of the
SAMs (solid lines) and ZCZ07 results for the ratio of central galaxy stellar 
mass at $z\sim 1$ and $z\sim 0$ as a function of present-day halo mass. 
Error bars in the SAMs predictions are the $1\sigma$ scatter around the mean. 
The dashed line (denoted as ``ZCZ07'') represents the standard ZCZ07 result. 
The dotted line (labeled as ``ZCZ07+'') is the conservative estimation, 
assuming a $25\%$ correction of the stellar mass in the DEEP2 sample 
(see text for details). }
\label{fig:Fig8MillHODerr}
\end{figure*}

Figure~\ref{fig:Fig8MillHODerr} shows the SAM predictions (solid lines) for 
the stellar mass as a function of halo mass and the results obtained by ZCZ07
(dashed lines), over the halo mass range they probe. 
Although the SAMs and ZCZ07 methods produce similar trends, there are 
important quantitative disagreements between them. The main difference is 
that the SAMs predict many more stars already in place at $z\sim 1$ in 
the progenitor central galaxies compared to the ZCZ07 results.
The differences are especially pronounced at medium and low halo masses. 
This may be related to the known fact that the SAMs produce too many 
$M_\star$ galaxies at high redshift (e.g., \citealt{Kitzbichler07}). 
At $z \sim 0$, on the other hand,  the agreement is quite good for low mass 
halos, while for high-mass halos (larger than $\sim 10^{12}\Msunh$) 
the SAMs seem to 
underpredict the stellar mass in central galaxies. The AGN ``radio mode'' 
feedback becomes important on these mass scales \citep{Bower06}.  This 
suggests that the SAMs might be overestimating the strength of the feedback. 

As for the fraction of stellar mass in place in the $z\sim 1$ progenitor
central galaxies, the SAMs prediction is about twice that inferred by ZCZ07, 
as shown in the bottom panels of Figure~\ref{fig:Fig8MillHODerr}. The error 
bars in the SAMs denote the $1\sigma$ scatter around the mean.
Note that this discrepancy could be less dramatic, as there might 
be $\sim 25\%$ underestimation in the ZCZ07 calculation of the stellar mass 
at $z\sim 1$ due to DEEP2 red galaxy incompleteness (see ZCZ07 for details). 
This effect is shown by the dotted lines in the bottom panels. Even with 
this potential correction, the discrepancy is significant. 

We note that these discrepancies are present at roughly the same level for
{\it both} SAMs, indicating that their cause is of a more
fundamental origin not reflected in the differences between the two
models. It is also worth mentioning that the major discrepancy between the 
SAMs and ZCZ07 predictions for the stellar mass growth appears to be
present  already at redshift 2. From the results presented
in \S3.4, we see that the amount of stars in place in central galaxies
by $z \sim 2$ predicted by the MPA SAM is of the order of the phenomenological
results for the amount of stars in place by $z \sim 1$.

\begin{figure*}
\includegraphics[width=165mm]{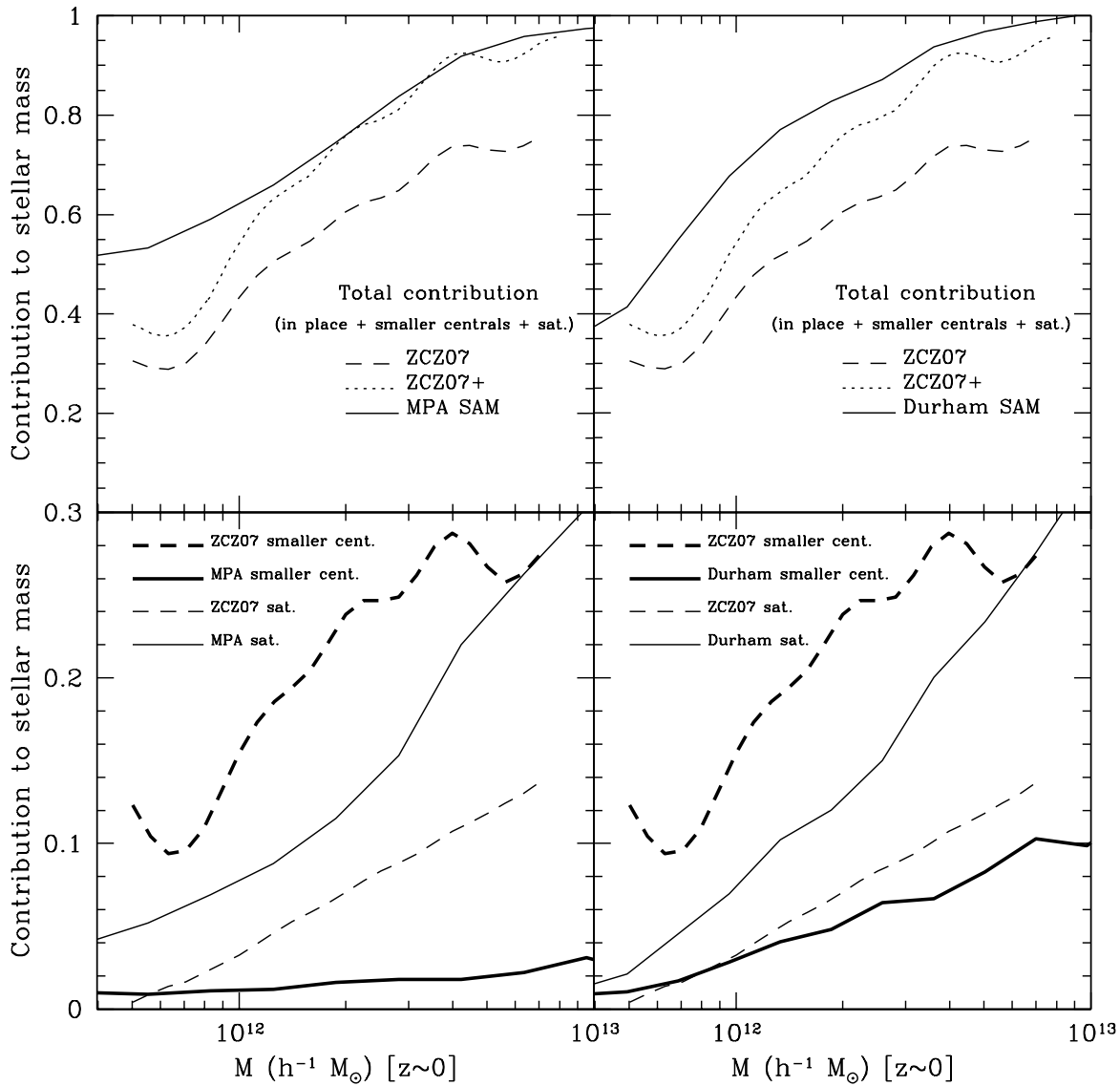}
\caption{{\it Top panels}: Predictions of the SAMs compared to the results 
obtained by ZCZ07 for the {\it total} stellar mass in central galaxies that 
was acquired through merging, in addition to that already in place at $z\sim1$.
The stellar mass is normalized by the amount of stellar mass at $z\sim0$ and
presented as a function of present-day halo mass. The 
left and right panels show the comparison to the MPA and the Durham SAMs, 
respectively (solid lines). The dashed lines are the standard ZCZ07 
result and the dotted lines denote the effect of a tentative 25\% 
correction of the stellar mass at $z \sim 1$ (see text for details). 
{\it Bottom panels}: Comparison between the SAMs (solid lines) and ZCZ07 
(dashed lines) for the {\it individual} contributions from mergers of 
smaller central galaxies (thick lines) and satellites (thin lines) to the 
stellar mass in the final central galaxy.}
\label{fig:Fig9MillHOD}
\end{figure*}

The top panels of Figure~\ref{fig:Fig9MillHOD} compare the predictions 
of the SAMs (solid lines) to the ZCZ07 results for the {\it total} stellar 
mass acquired through merging of smaller central and satellite galaxies on 
top of that already in place at $z\sim 1$, normalized by the final stellar 
mass at $z\sim 0$. 
For ZCZ07, we plot both the standard estimation (dashed line in each panel)
and the conservative estimate including the possible 
25\% correction of the stellar mass at $z \sim 1$ (dotted line). 
Here the agreement is better, especially at high halo masses.
The difference between these total contributions 
and the stellar mass at present-day in halos of a given mass is essentially 
the contribution arising from star formation since $z\sim1$.

Finally,
the bottom panels of Figure~\ref{fig:Fig9MillHOD} contrast the individual
contributions to the stellar mass assembly in the SAMs (solid lines) 
and ZCZ07 (dashed lines),  showing the relative   
contributions to the final central galaxy from mergers of smaller central 
galaxies (thick lines) and satellite mergers (thin lines).
The individual contributions from these models are strikingly different. 
In ZCZ07, the main merger contribution to present-day central galaxies comes from 
mergers of the smaller central progenitor galaxies, while in the SAMs a
larger contribution comes from satellite galaxies. 
As mentioned already in \S3.3, the contribution from smaller central 
progenitor galaxies in the MPA model is particularly minor, while it is 
somewhat larger in the Durham model. The implications of these differences 
are discussed further below.

\subsection{Examining Assumptions in the ZCZ07 Approach}

In this section we use the SAM catalogs to test the validity of some of the 
assumptions adopted in the ZCZ07 method. We note that these tests 
can be done despite the discrepancies found between the SAMs and ZCZ07.

In ZCZ07, an evolutionary link is established between galaxy populations 
at the two epochs via the theoretically predicted growth of dark matter halos
in which these galaxies reside.  The first assumption we 
examine is related to the statistical nature of any such relation.
ZCZ07 use the average relation between masses of present-day halos and 
that of their main progenitors at $z\sim1$, obtained by the PINOCCHIO 
code \citep{Monaco02}, with no accounting for the intrinsic scatter. 
This could, in principle, impact the estimation of the stellar mass growth. 

Using the SAMs, where the full merger tree is known, it is possible to 
test this.
For each individual $z\sim 0$ central galaxy, the SAM catalog provides its 
present-day stellar mass and host halo mass as well as its main progenitor 
stellar mass and host halo mass at $z\sim 1$.
Based on these, we derive the the stellar mass of central 
galaxies and that of the progenitor central galaxies as a function of 
the present-day halo mass, which by construction includes the scatter in the 
halo mass and progenitor halo mass relation. With the SAM catalog, we can 
also follow the ZCZ07 procedure to obtain the above relations by using the 
{\it average} growth of dark matter halos. 

\begin{figure}
\includegraphics[width=\columnwidth]{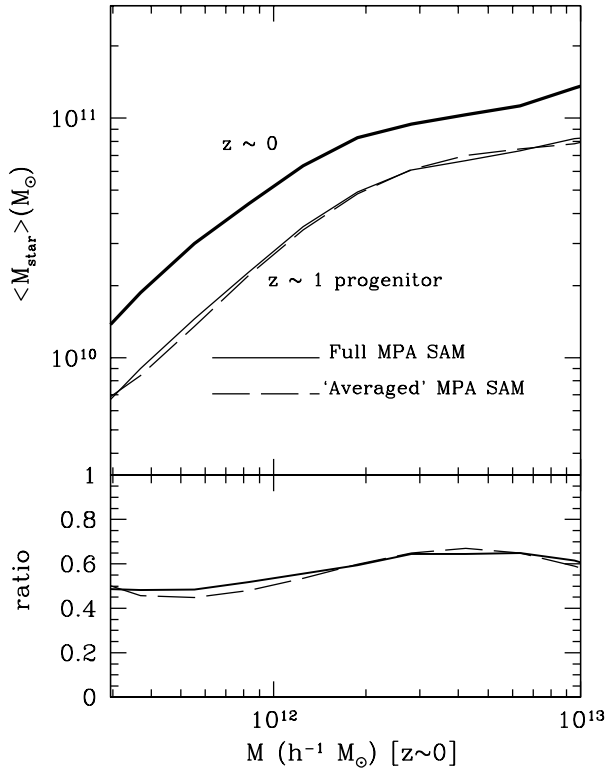}
\caption{The MPA predictions for the mean stellar mass in central galaxies at 
$z \sim 0$  and that of their $z \sim 1$ main progenitors as a function of 
the present-day halo mass, calculated using the full halo growth information 
(solid lines) and the ``averaged'' relation from the ZCZ07 approach (dashed lines). See 
text for details.}
\label{fig:Fig8FK}
\end{figure}

We perform such a test with  the MPA catalog.
Figure~\ref{fig:Fig8FK} compares the stellar mass evolution as a function of 
present-day halo mass obtained with those two different procedures. The 
solid lines are the results obtained using the individual halo growth
information (presented already by the solid lines in 
Fig.~\ref{fig:Fig8MPADurham}), while the dashed lines denote the 
predictions using the {\it average} halo growth.  
There are only negligible differences between both results, with a 
maximum deviation of about 5\%, indicating that the use of the average
halo growth to connect galaxies at the two epochs is adequate.

The other assumptions in ZCZ07 are related to the rough estimation of 
the merger contributions to the stellar mass growth of central galaxies,
resulting in the dashed curves shown in the bottom panels of
Figure~\ref{fig:Fig9MillHOD}. Computing these contributions 
in a statistical way is not straightforward. 
In order to estimate the stellar mass contribution from smaller central
galaxies, ZCZ07 utilize, for a given dark matter halo mass,
the fraction of halo mass already formed at $z \sim 1$ in its 
main progenitor. This halo has some fraction of stars already formed by 
that epoch. The assumption is that the ratio of central galaxy stellar mass to 
halo mass in the other progenitors is the same as that for the main progenitor
and that they all will have merged into the main progenitor central galaxy 
by the present time. For example,  a halo of mass $5 \times 10^{11} \Msunh$ 
(at present) has assembled about  $70\%$ of its mass 
by $z \sim 1$, while the MPA SAM predicts that the corresponding central galaxy
has about $50\%$ of the stars already in place by $z \sim 1$. Following 
the ZCZ07 procedure, the remaining $30\%$ of the halo should  contribute 
$30/70 \times 50\% \sim 21\%$ of the stars. Implicitly, this computation 
assumes a constant star formation efficiency (or that the merging halos 
are of comparable mass). 
Obtaining the contribution of satellite galaxies to the final central 
galaxies is more difficult. ZCZ07 provide a crude estimation based on 
simplifying assumptions resulting in a linear extrapolation from zero at
the low-mass end, where the satellite contribution is expected to be
negligible, to  $25\%$ of the central galaxies contribution at the 
high-mass end, where the brightest satellite in each halo is assumed to 
have merged with the central galaxy (see their \S6.3 for more details).

\begin{figure*}
\includegraphics[width=165mm]{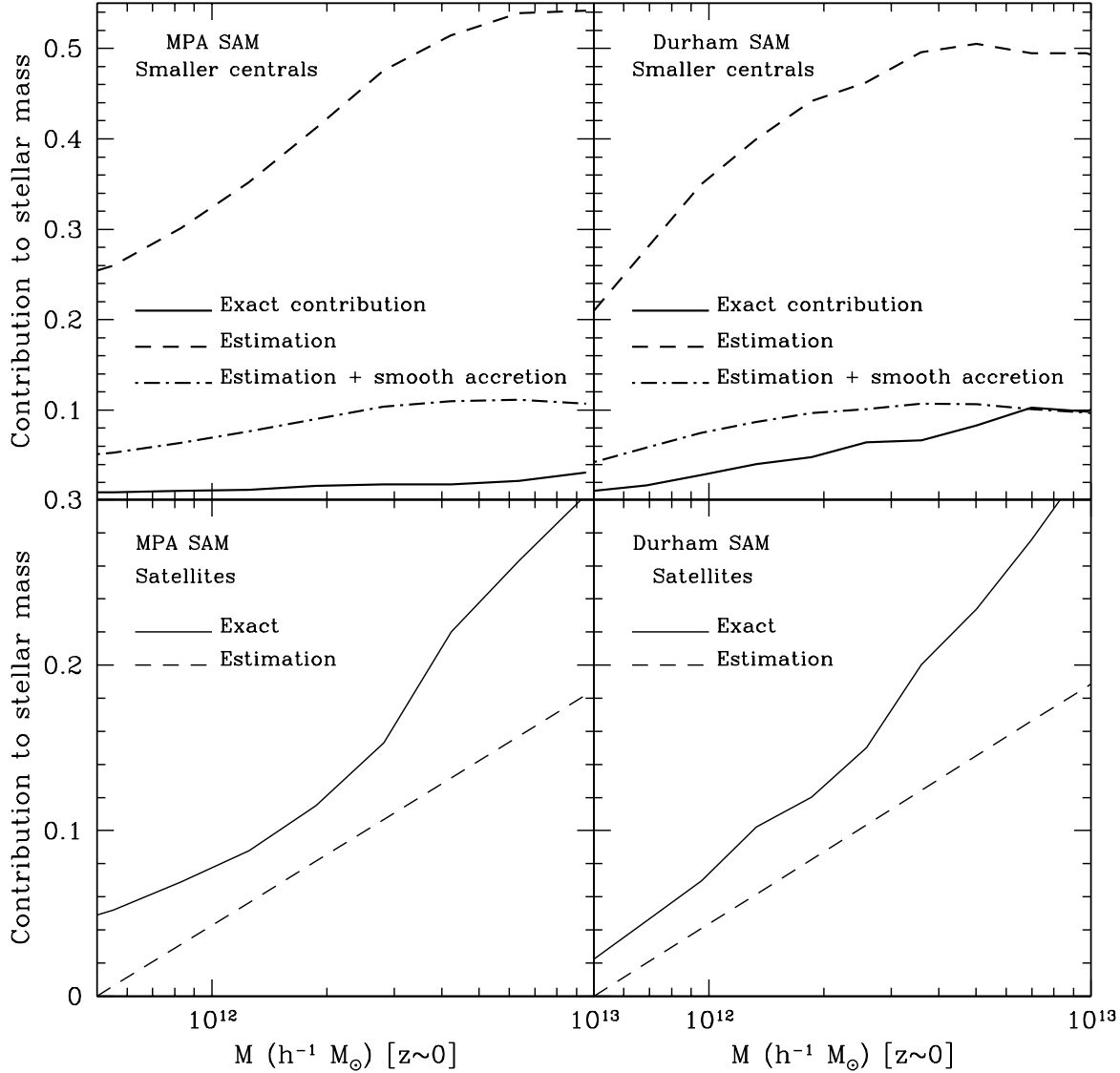}
\caption[]{
\label{fig:contTest}
Testing ZCZ07 assumptions regarding the contribution to the central
galaxy stellar mass from mergers of smaller central galaxies and satellites 
since $z\sim 1$ with the MPA SAM (left) and Durham SAM (right).  The 
contribution is normalized by the total amount of stellar mass at $z\sim 0$ 
and presented as a function of present-day halo mass. 
{\it Top panels:} The solid line in each panel denotes the ``exact'' stellar 
mass contribution from the merging of smaller central galaxies. The 
predictions for 
this contribution using the ZCZ07 assumptions applied to the SAMs are shown
as the dashed-lines (marked as ``Estimation''). The dot-dashed lines 
incorporate the effect of ``smooth accretion'' (see text).
{\it Bottom panels:} The same now for the stellar mass contribution from 
merging satellites. In each panel, the solid line shows the exact contribution 
from mergers of satellites into the final central galaxy, while the dashed 
line is the estimated contribution using the ZCZ07 approximation.
} 
\end{figure*}

We use the full halo and galaxy merging histories provided in the SAMs to 
test these assumptions regarding the contribution from smaller central 
galaxies and from satellites. The top panels of Figure~\ref{fig:contTest} 
show for both SAMs the predicted contribution of smaller 
central galaxies to the final stellar mass obtained by applying the simple 
estimation proposed by ZCZ07 (dashed lines). The exact contribution in 
the SAMs is shown by the solid lines. It is apparent that the approximate
approach overestimates the ``real'' contribution by a large amount.
Recall that the SAMs appear to predict a large excess of 
stars that already assembled in the central progenitors at $z \sim 1$
(Fig.~\ref{fig:Fig8MillHODerr}), which might translate to an overestimation
of the estimated contribution.

Another important issue is that the estimation assumes that 
all the remaining halo mass accreted carries with it the same fraction 
of stellar mass. This may be reasonable for the merging smaller halos,
however, there is also a significant contribution from 
smooth ``diffuse'' accretion of dark matter particles to the final halo.  
In the Millennium Simulation, we obtain that this component accounts for 
$\sim 30\%$ of the final halo mass since $z\sim 1$ (see also 
\citealt{Guo08,Fakhouri09,Genel10}). Even 
though smooth accretion carries baryons that will be added to the hot 
gas available for cooling \citep{DeLucia04}, it does not contribute to the 
existing stellar mass. When accounting for this, we find that 
the contribution from smaller central galaxies is considerably reduced (the 
dot-dashed lines in 
top panels of Fig.~\ref{fig:contTest}) and is in better agreement with the 
SAMs predictions (particularly for the Durham model).  This approximated 
contribution, however,  still appears to be somewhat overestimated, 
especially in the MPA model, where the measured contribution of smaller 
centrals is tiny. 

As mentioned already, the simplified estimate also implicitly 
assumes a constant SFE 
for the merging halos, which in reality does vary with halo mass (see, e.g., 
Fig.~\ref{fig:Fig10mill50b} in this work and Fig.~10 in ZCZ07).
This can affect the contribution from smaller central galaxies in two ways: 
for a halo whose mass is around the peak of SFE or smaller, this assumption
might overestimate the contribution, while for larger mass halos 
it can result in an underestimation (which will result in a stronger 
downsizing behavior). 

The bottom panels of Figure~\ref{fig:contTest} compare
the rough estimation of the merger contribution from satellites to the 
stellar mass of the central galaxy (dashed lines) with the SAMs
exact predictions (solid lines). The estimation is computed in the
same halo mass range used in ZCZ07, to facilitate the comparison. In
this case, we see that the approximation underestimates the merger 
contribution 
of satellites to the final stellar mass, especially at the highest halo 
masses probed. This may arise due to the merging of additional satellites
to the brightest one in each progenitor halo, and may again lead to a 
stronger downsizing effect.
We note that the behavior of the estimated satellite merger
contribution goes in the opposite direction than that of 
smaller central galaxies. This leads to a partial cancellation such
that the estimate of the total merger contribution is reasonable, which
also implies that the inferred contribution from recent star formation is not
strongly affected by the approximate nature of the above method.

It is hard to draw definitive conclusions from all these tests given the
intrinsic uncertainties in both the SAMs and the ZCZ07 estimations. We 
clarify that the latter assumptions investigated here have to do with
transforming the ZCZ07 measurement of the growth of stellar mass in central
galaxies as a function of halo mass (their Fig.~8), which is 
robust, to the overall contribution of mergers and star formation to
the stellar mass assembly (their Fig.~9). More care should certainly be
given to these assumptions, taking into account smooth accretion and
incorporating better treatment of central and 
satellite dynamics determined from analytic models (e.g., \citealt{Zentner05})
or from simulations (e.g., \citealt{wang06,White07,Wake08}). Nonetheless,
we expect the qualitative results of ZCZ07 to still be valid (or even
somewhat strengthened with these corrections, as discussed above), and 
believe that such phenomenological methods can provide powerful constraints
on theories of galaxy formation and evolution.

\section{Summary and Discussion}
In this paper we study theoretical predictions for the evolution of
stellar mass in galaxies as a function of their host halo mass, using
semi-analytic galaxy formation models based on the Millennium simulation.
We utilize two different SAM implementations, the MPA 
\citep{Croton06,DeLucia07} and Durham \citep{Bower06} models.
We investigate the different contributions to the growth of stellar
mass, and the role of mergers and star formation in the
stellar mass assembly from $z\sim1$ to the present. 
Such an investigation with SAMs is timely and important as
several recent studies have started to explore the galaxy-halo
connection and related inferences on galaxy evolution (e.g., ZCZ07; 
\citealt{White07,Conroy09,Wake10,wang09,Behroozi10,Firmani10,Guo10,Leauthaud11,Neistein11}). 
These studies employ different
phenomenological approaches utilizing observed statistical
properties of galaxies, such as correlation functions, abundances
of galaxies and stellar mass functions. In our study,
we particularly compare the SAM predictions to the methodology
and results presented in ZCZ07, to assess the potential of such studies
to constrain galaxy formation models and to guide future efforts of 
modeling galaxy evolution.

We find that the SFE, the fraction of baryon mass that has converted into
stars in the central galaxies, 
as a function of halo mass has a peaked distribution, with a maximal
value of $\sim 23\%$ at $z\sim0$ and $\sim18\%$ at $z\sim1$. The location
of the peak shifts toward lower halo mass with time, reflecting 
``halo downsizing''. 
Both SAM models produce similar results for the growth of stellar mass
in central galaxies from $z\sim1$ to 0 as a function of the present-day
halo mass.  At both redshifts, the central galaxy stellar mass increases
rapidly with halo mass for relatively low-mass halos (below  $\sim 2 \times 
10^{12} \Msunh$ ) and at a lower rate for more massive halos. The fraction
of stellar mass already in place at $z\sim1$ also varies with final
halo mass:
it is about $50\%$ ($40\%$) for the MPA (Durham) model at the low-mass end,
increasing with halo mass to about $65\%$ for halos with mass 
$\sim$ a few $\times 10^{12} \Msunh$, and decreases somewhat at the
highest halo mass probed. 

The SAM predictions for the different contributions to the stellar
mass assembly since $z\sim1$ indicate that star formation 
is more important in low mass halos 
($\sim$ a few $\times 10^{11} \Msunh$), 
while accretion through mergers dominates at the high-mass end 
($\sim 10^{13} \Msunh$) , where star formation is negligible.
In the intermediate regime both these processes contribute.
This trend with halo mass is another manifestation of downsizing.
The two SAMs provide similar results, differing mostly in
their predictions for the contribution of smaller central galaxies merging
with the main central galaxy. This likely arises from differences 
in the galaxy formation prescriptions and in the merger 
trees of these models.

We also study the predictions of the MPA SAM for the stellar mass growth 
since $z \sim 2$. The trends found are very similar to those for $z \sim 1$, 
including the presence of the ``downsizing'' pattern. As expected, much less 
stellar mass is already in place in the main progenitors compared to $z\sim 1$ 
and the contribution from merging of smaller central galaxies is considerably 
larger. Furthermore, the contribution from star formation is 
important at all halo masses, even at the high-mass end.

Our study is motivated by ZCZ07 who develop a 
novel phenomenological approach to study galaxy evolution by connecting
galaxy clustering results at different epochs through the growth of
the hosting halo mass.  Such applications can potentially provide important
constraints  for galaxy formation models as a function of the host halo
mass, which is a fundamental parameter in such models.
We compare our finding to those of ZCZ07.
We find that the SAMs and ZCZ07 produce similar trends for the stellar mass
assembly in halos, however, there are significant quantitative differences.
The SFE of central galaxies as a function of halo mass at both epochs 
in the SAMs and ZCZ07 are qualitatively similar, with the same overall
peaked shape and  halo downsizing. The main discrepancy appears at $z \sim 1$ 
where the MPA SAM predicts a $\sim 50\%$ higher SFE than ZCZ07. 

The differences are also apparent when contrasting, for these two approaches,
the stellar mass content of halos at the two epochs as a function of 
present-day halo mass (Fig.~5). While the overall trends are in 
qualitative agreement, there are striking differences.
Again, the most significant difference is that the SAMs predict a larger
stellar mass content at $z \sim1$ for all halo masses (with a bigger 
discrepancy for lower-mass halos). The results from the SAMs and ZCZ07 are 
in better agreement at $z \sim 0$, however,  the SAMs 
underestimate the stellar mass in central galaxies in present-day halos more
massive than $\sim 10^{12} \Msunh$.
Note that for halos larger than this characteristic mass, AGN feedback 
starts to playing an important role (see e.g., \citealt{Stringer09}), 
suggesting its effect might be overestimated within these models,
resulting in over-quenching stellar mass growth. 

When looking at the fraction of stellar mass in 
present-day central galaxies that is already in place at $z \sim 1$,
there is a factor of two disagreement between the SAMs and the ZCZ07 
results. For example, 
ZCZ07 obtain that about $30\%$ of the stars in halos of $\sim 3 
\times 10^{12} \Msunh$ are formed by $z \sim 1$, while the SAMs predict 
about $60\%$.  The discrepancy is significant even when conservatively
accounting for a possible underestimation of the DEEP2 stellar masses.
Note also that these differences are already present at higher redshifts, 
as we find that the ratio of stellar mass in central galaxies at $z \sim 2$ 
in the MPA SAM is comparable to the one predicted by ZCZ07 at $z \sim 1$. 
Our results are in agreement with previous works that studied SAM predictions, 
which found an excess of stars already in place by $z \sim 1$ (e.g., 
\citealt{Croton06, Kitzbichler07}). Those, however, were focused on integrated 
properties and not explicitly as a function of halo mass as we  show here. 

The SAM predictions for the total amount of stars acquired through merging
on top of that already in place at 
$z \sim 1$ are, at first order, in good agreement with ZCZ07 results. However, 
the individual contributions to the central galaxies stellar mass 
from mergers of smaller centrals and satellite mergers are markedly
different. 
In ZCZ07 the significant merger contribution arises from the smaller 
central progenitors. In contrast, the SAMs predict a large contribution
from mergers of satellites.
It is the partial cancellation of these opposing differences that leads
to a reasonable agreement of the total merger contribution. As a whole, 
the SAMs and the ZCZ07 results lead to a similar behavior of the star 
formation contribution with halo mass.

While the comparison between the ZCZ07 observational results and the SAMs 
predictions is informative, there are some simplified assumptions in the 
former. ZCZ07 apply the method as a proof of concept and point out that 
there is room for improvement with more sophisticated applications. 
With the SAMs, we are able to examine different working assumptions 
employed by ZCZ07 and guide these efforts. 
For instance, to link galaxies at two epochs, ZCZ07 use the average 
relationship between the present-day halo mass and the mass of the main 
progenitor at $z \sim1$, neglecting the individual scatter among halos.
We test the validity of this assumption, using the full assembly information
available in the SAMs, finding that it results in negligible differences.

On the other hand, some of the assumptions made by ZCZ07 to estimate the 
overall contribution from mergers and star formation can certainly be
improved.  In particular, the 
original ZCZ07 estimation does not take into account the smooth accretion of 
dark matter particles to the final halo mass. 
In the Millennium Simulation the smooth accretion since $z \sim1$ accounts 
for about $\sim 30\%$ of the final halo mass. 
It is difficult, however, to estimate the contribution of stellar mass 
from very small halos, below the resolution of current numerical simulations.
If smooth accretion is as significant in the real universe, 
it will certainly be needed to be included in such approximations.
More realistic SFE dependence on halo mass can also be implemented to
improve the estimation method.

Conroy \& Wechsler (2009) present related calculations for the evolution 
of stellar masses and star formation. They use abundance matching to  
monotonically link galaxies to halos. They predict similar, but more 
dramatic trends than both the SAMs and the ZCZ07 approach. For instance, 
they suggest that galaxies in lower mass halos ($\sim 10^{11}\Msunh$) grow 
their stellar mass purely by star formation, while essentially all of the 
mass is already in place by $z \sim 1$ in present-day halos of $10^{13} 
\Msunh$. Their results support a picture in which stellar mass grows only
via star formation,  suggesting that the stellar mass 
from smaller progenitors does not merge into the central galaxy, but remains 
as satellites or diffuse light.
Additional studies are needed to fully clarify their differences with the
results presented here and in ZCZ07.
Recently, \citet{Neistein10} has critically studied the assumptions made 
in the abundance matching method using SAM catalogs and found important 
differences, indicating that environmental processes may be important.

We have demonstrated that phenomenological methods such as ZCZ07 are
powerful for studying galaxy formation and evolution. They provide key 
constraints for theoretical models,  such as the SAMs, as a function of 
halo mass. By highlighting
remaining shortcoming of galaxy formation models, they can guide to 
improving theoretical predictions at high redshift and increase 
our understanding of the complex picture of galaxy formation and evolution.
At the same time, while ZCZ07 is useful as  a proof of concept, 
future work should use more sophisticated methods applied to better data,
and the SAMs can serve an important role in developing and testing
such methods.

Future work will also explore the role of environment in the buildup of stellar
mass within the host halos. One of the main assumptions in the current HOD 
framework is that the galaxy content in halos depends only on the halo 
mass, and is independent of the large-scale environment where the halo is 
located. Recent theoretical studies have shown that the clustering
properties of dark matter halos depend on the large-scale environment
(the so-called halo assembly bias; \citealt{Gao05,Wechsler06,Croton07,Jing07}).
This environmental effect might also impact galaxy properties and galaxy 
clustering 
(\citealt{Zhu06, Zu08}). Using the SAMs, we may be able to test the effect of 
large-scale environment on stellar mass assembly and galaxy evolution 
(c.f., \citealt{Hoyle11}). Moreover, we could use SAM 
results to incorporate environmental effects to phenomenological methods 
such as ZCZ07, increasing the constraining power of galaxy 
clustering data on galaxy formation models.

\acknowledgments

We thank Gabriella De Lucia and Gerard Lemson for providing 
substantial help with using the Millennium Databases.
This work has been supported by NSF grant AST-0907947.
I.Z.\ was further supported by NASA through a contract issued
by the Jet Propulsion Laboratory. 
Z.Z.\ gratefully acknowledges support from the Yale Center for 
Astronomy and Astrophysics through a YCAA fellowship.
The Millennium Run simulation used in this paper was carried out by the
Virgo Supercomputing Consortium at the Computing Center of the Max-Planck
Society in Garching. The web application providing on-line access to them 
were constructed as part of the activities of the German Astrophysical 
Virtual Observatory.


\end{document}